\documentclass[lineno,authoryear]{FLO_v1_preprint}%

\usepackage{graphicx}
\usepackage{upgreek}
\usepackage{multicol,multirow}
\usepackage{amsmath,amssymb,amsfonts}
\usepackage{mathrsfs}
\usepackage{amsthm}
\usepackage[figuresright]{rotating}
\usepackage{appendix}
\usepackage[authoryear]{natbib}
\usepackage{ifpdf}
\usepackage[T1]{fontenc}
\usepackage{newtxtext}
\usepackage{newtxmath}
\usepackage{textcomp}
\usepackage{xcolor}

\usepackage[colorlinks,allcolors=blue]{hyperref}
\definecolor{jourcolor}{cmyk}{1,0.57,0.01,0.38}
\hypersetup{
    colorlinks,%
    citecolor=jourcolor,%
    filecolor=jourcolor,%
    linkcolor=jourcolor,%
    urlcolor=jourcolor
}

\theoremstyle{definition}

\articletype{PREPRINT}

\begin{document}

\title[]{Wind speed inference from environmental flow-structure interactions, part 2: leveraging unsteady kinematics}

\author[]{Jennifer L. Cardona$^{1}$ and John O. Dabiri$^{2\ast}$}



\address[1]{\orgdiv{Department of Mechanical Engineering}, \orgname{Stanford University}, \orgaddress{\street{Stanford}, \state{California}, \postcode{94305}, \country{USA}}}

\address*[2]{\orgdiv{Graduate Aerospace Laboratories \& Mechanical Engineering}, \orgname{California Institute of Technology}, \orgaddress{\street{Pasadena}, \state{California}, \postcode{91125}, \country{USA}}}


\corres{*}{Corresponding author. E-mail:
\emaillink{jodabiri@caltech.edu}}



\abstract{This work explores the relationship between wind speed and time-dependent structural motion response as a means of leveraging the rich information visible in flow-structure interactions for anemometry. We build on recent work by \citet{Cardona2021WindInteractions}, which presented an approach using mean structural bending. Here we present the amplitude of the dynamic structural sway as an alternative signal that can be used when mean bending is small or inconvenient to measure. A force balance relating the instantaneous loading and instantaneous deflection yields a relationship between the incident wind speed and the amplitude of structural sway. This physical model is applied to two field datasets comprising 13 trees of 4 different species exposed to ambient wind conditions. Model generalization to the diverse test structures is achieved through normalization with respect to a reference condition. The model agrees well with experimental measurements of the local wind speed, suggesting that tree sway amplitude can be used as an indirect measurement of mean wind speed, and is applicable to a broad variety of diverse trees.}

\maketitle

\begin{boxtext}

\textbf{\mathversion{bold}Impact Statement}

It has recently been proposed that environmental structures such as trees can be used as ubiquitous, low-cost flow sensors by leveraging visual observations of their characteristic responses to wind loading \citep{Cardona2021WindInteractions}. Potential application areas include analyses of pollution dispersal and wildfire propagation. The present work demonstrates that measurements of tree sway amplitudes can be related to wind speeds in the context of this visual anemometry goal. This greatly expands the potential of visual anemometry methods to be used on a broad variety of trees and in lower-speed wind conditions, since it does not require that trees exhibit large observable mean bending.
\end{boxtext}

\section{Introduction}
\label{sec:intro}
Recent work has suggested that flow speeds can be measured using visual observations of flow-structure interactions such as the bending response of trees to incident wind \citep{Cardona2019SeeingNetwork, Cardona2021WindInteractions}. Video of trees could potentially be used instead of conventional point-wise anemometers to achieve low cost, spatially resolved wind speed measurements simultaneously throughout the entire camera field of view. Such data would be valuable for applications such as analyses of pollution dispersal and wildfire propagation. \citet{Cardona2021WindInteractions} presented a technique that allowed for normalized wind speed measurements to be made using mean structural deflections compared to a known reference condition. The aforementioned method demonstrated how visual observations of objects including trees can be used toward anemometry tasks. However, the application of that method requires that the mean deflections can be observed, which may not always be possible or convenient in practice. Mean deflections may be small in many cases, especially for large trees, making them difficult to measure accurately. For instance, a field study by \citet{Peltola1993SwayingMeasurements} found that for two Scots pines of 9.5 and 13.5 m tall, the maximum bending measurements at the crown centre heights were less than 2 and 5 cm respectively for wind speed ranges of 4-8 ms$^{-1}$. The amount of static bending that a tree undergoes is a function of its slenderness and material properties in addition to the wind (discussed further in section \ref{sec:regimes}). This limits the range of trees and wind conditions for which the mean bending method proposed by \cite{Cardona2021WindInteractions} can be used. Another challenge is that mean deflections must be measured with respect to the object position under no wind load. This inherently means that the object must be observed in the absence of wind in addition to being observed under a known calibration wind speed.


Trees are desirable target objects to use for visual anemometry because of their ubiquity. For instance, in New York City, more than $680,000$ trees have been mapped to date, with 234 species represented \citep{NYCParks2021NewMap}. A tree-based visual anemometry method will be most widely applicable to wind mapping applications if it can be used on a diverse range of trees of various sizes and species. Although the mean deflection-based technique developed by \citet{Cardona2021WindInteractions} is limited in this regard, the behavior of a tree in response to incident wind is dynamic, and is rich with information extending beyond the mean deflections. For example, the land adaptation of the Beaufort scale relies on perceptible tree branch motion as a correlate for low wind speeds \citep{Jemison1934BeaufortMountains}. Tree motion has previously been related to the time-varying wind speed, often through semi-empirical mechanical transfer functions \citep{Mayer1987, Holbo1980, Kerzenmacher1998AWind, Moore2008SimulatingMethod}. These transfer functions depend on the tree-specific properties affecting the dynamics, including the natural frequency, damping, and drag coefficient. Prior work measuring tree sway also suggests that the magnitude of the tree sway increases with increasing wind speed \citep{Peltola1993SwayingMeasurements, vanEmmerik2017MeasuringAccelerometers}, as does the velocity of the tree branches \citep{Tadrist2018}.


The present work leverages dynamic tree sway to extend the capabilities of the visual anemometry technique proposed by \citep{Cardona2021WindInteractions}. A physical model is developed, applying a force balance to relate mean wind speeds to the amplitude of tree sway. This eliminates the need to observe the object of interest in the absence of wind loading, and allows measurements to be made in cases where swaying behavior is observable even when mean deflections are difficult to measure. The model is applied to two field datasets including a video dataset capturing the swaying behavior of a \textit{Magnolia grandiflora} in an open field, and a publicly available dataset with strain gage measurements of 12 trees of 3 different species in a broadleaf forest in the United Kingdom \citep{Jackson2018StrainUK}. Results suggest that the method developed in this work is widely applicable to a diverse set of trees.

\section{Methods}
\subsection{Analytical model}
\label{sec:derivation}

The structural deflection, $\delta$, in response to wind loading was modelled based on Newton's law for a single degree of freedom damped harmonic oscillator:

\begin{align}
    F_W(t) = m \ddot{\delta} + \lambda \dot{\delta} + \kappa \delta
    \label{eq:oscillator}
\end{align}

\noindent where $F_W$ is the external forcing due to incident wind, $m$ is the mass of the structure, $\lambda$ is the damping coefficient, and $\kappa$ is the elastic constant. An arbitrary time-dependent forcing $F(t)$ can be expressed as a Fourier series:

\begin{align}
F(t) &= \frac{a_0}{2} + \sum^{\infty}_{n=1} \left( a_n \cos(nt) \right) + \sum^{\infty}_{n=1} \left( b_n \sin(nt) \right)
\end{align}

When forced with harmonic loading:

\begin{align}
    F(t) = F_0 \sin(\omega t)
\end{align}

\noindent where $\omega$ is the forcing frequency, and $t$ is time, the steady-state solution of equation \ref{eq:oscillator} is:

\begin{align}
    \delta(t) = \frac{F_0}{\kappa} \left[ \frac{1}{(1-\beta^2) + (2 \zeta \beta)^2}  \right] \left[ (1-\beta^2) \sin(\omega t) - 2\zeta \beta \cos(\omega t)  \right]
    \label{eq:ss_solution}
\end{align}

\noindent where $\beta$ is the ratio of the forcing frequency to the natural frequency of the system $(\beta = \omega/\omega_n)$, and $\zeta = \frac{\lambda}{2\sqrt{\kappa m}}$ \citep{Clough1995DynamicsStructures}.


In the present work, we define the amplitude of the structural oscillations as the standard deviation of the structural deflection, $\sigma(\delta)$ (median absolute deviation is discussed as an alternative to this choice in the supplementary material, section \ref{sec:mad}). The steady-state solution given by equation \ref{eq:ss_solution} reveals that the amplitude of structural oscillations scales with the forcing amplitude:

\begin{align}
    \sigma(\delta(t)) \propto \sigma(F_W(t)) \label{eq:delta_to_F} 
\end{align}

The instantaneous force of the wind on the structure, $F_W$, is given by:

\begin{align}
    F_W(t) &\propto \rho A U^2(t) \label{eqn2}\\
    &= CU^2(t) \label{eq:FW_prop}
\end{align}

\noindent where $\rho$ is the fluid density, $A$ is the projected frontal area of the structure, $U(t)$ is the instantaneous wind speed, and $C$ is a positive constant. The instantaneous wind speed, $U(t)$, can be decomposed as a sum of the mean wind speed, $\bar{U}$, and the unsteady fluctuating wind speed, $u'(t)$. This gives:

\begin{align}
   F_W(t) &=C [\bar{U} + u'(t)]^2 \nonumber\\
    &=C [\bar{U}^2 + 2\bar{U}u' + u'^2]  \label{eq:FW_to_U}
\end{align}

\noindent Taking the standard deviation of equation \ref{eq:FW_to_U} gives an expression relating $\sigma(F_W)$ to $\bar{U}$ and $u'$:

\begin{align}
    \sigma(F_W) &= \sigma \left( C [\bar{U}^2 + 2\bar{U}u' + u'^2] \right ) \nonumber\\
    &= C \sigma \left( \bar{U}^2 + 2\bar{U}u' + u'^2 \right ) \nonumber\\
    &= C \sigma \left(2\bar{U}u' + u'^2 \right ) \label{eqn_above}
\end{align}

\noindent or equivalently:

\begin{align}
\sigma(F_W) &\propto \sigma(2\bar{U}u' + u'^2)
\label{eq:std_delta}
\end{align}

Assuming $u'/\bar{U} << 1$, the $u'^2$ term in equation \ref{eq:std_delta} can be neglected, giving:

\begin{align}
    \sigma(F_W) &\propto \sigma( 2\bar{U}u') \nonumber\\
    &\propto \sigma(\bar{U}u') \nonumber\\
    &\propto \sigma(u')\bar{U} \nonumber\\
    &\propto \frac{\sigma(u')}{\bar{U}} \bar{U}^2 \label{eq:F_final}
\end{align}

\noindent The assumption that $u'/\bar{U} << 1$ is examined in field measurements described below. The turbulence intensity, $\frac{\sigma(u')}{\bar{U}}$, will be denoted as $I_u$. Given that that the structural sway amplitude, $\sigma(\delta)$, scales with $\sigma(F_W)$ (equation \ref{eq:delta_to_F}), $\sigma(\delta)$ can replace $\sigma(F_W)$ in equation \ref{eq:F_final}, yielding:  

\begin{align}
    \sigma(\delta) &\propto I_u \bar{U}^2 \label{eq:dynamic_model}
\end{align}

\noindent The simplified final expression given in equation \ref{eq:dynamic_model} reveals that the sway amplitude scales with $\bar{U}^2$ multiplied by the turbulence intensity $I_u$.

\begin{table*}[!h]
  \begin{center}
\caption{ Summary of the properties of the 13 test trees analyzed in the present work, including species, height ($h$), diameter at breast height (DBH), and approximate elastic modulus ($E$). Literature-reported values of $E$ were obtained from $^{a}$\citet{Green1999MechanicalWood} and $^{b}$\citet{Niklas2010WorldwideDensity}. Tree ID numbers assigned in the original \citet{Jackson2018StrainUK} dataset are also listed for the relevant subset of trees.}
\vspace{12 pt}
  \begin{tabular}{lcccccc}
      Dataset & Tree ID \# & Species & $h$ (m) & DBH (m) & $E$ (GPa)\\
     \midrule
      Present & N/A & Magnolia & 5 & 0.08 & 9.7$^a$\\
      \citet{Jackson2018StrainUK} & 8 &  Ash & 23.37 & 0.26& 9.5$^b$\\
      \citet{Jackson2018StrainUK} &9 & Ash & 24.37 & 0.34& 9.5$^b$\\
      \citet{Jackson2018StrainUK} &10 & Ash & 23.87 & 0.28& 9.5$^b$\\
      \citet{Jackson2018StrainUK} &11 & Ash & 18.91 & 0.24& 9.5$^b$\\
      \citet{Jackson2018StrainUK} &13 & Ash & 22.10 & 0.37& 9.5$^b$\\
      \citet{Jackson2018StrainUK} &14 & Ash & 23.40 & 0.38& 9.5$^b$\\
      \citet{Jackson2018StrainUK} &15 & Sycamore & 18.41 & 0.23& 8.4$^b$\\
      \citet{Jackson2018StrainUK} &17 & Ash & 23.20 & 0.39& 9.5$^b$\\
      \citet{Jackson2018StrainUK} &18 & Birch & 16.28 &0.15& 9.9$^b$\\
      \citet{Jackson2018StrainUK} &19 & Birch & 16.07 & 0.24& 9.9$^b$\\
      \citet{Jackson2018StrainUK} &20 & Birch & 15.34 & 0.21& 9.9$^b$\\
      \citet{Jackson2018StrainUK} &21 & Birch & 19.90 & 0.28& 9.9$^b$\\
      
  \end{tabular}
  \label{tab:jackson_trees}
  \end{center}
\end{table*}

\subsection{Field Measurements}
\label{sec:experimental_methods}

Analysis was carried out on two distinct datasets: a direct field measurement dataset comprising videos of a \textit{Magnolia grandiflora} collected for the purposes of this work, and an indirect field measurement dataset comprising strain gage data for several trees collected by \citet{Jackson2018StrainUK}. Methods are described for each of these two datasets in sections \ref{sec:flora_methods} and \ref{sec:sg_methods} respectively. The direct field measurements from the videos of the \textit{Magnolia grandiflora} are used to demonstrate model application to visual observations of dynamic oscillations. Model generalizability to a diverse set of trees is established through its application to the indirect field measurements from \citet{Jackson2018StrainUK}.

In total, 13 trees across 4 different species were analyzed between the two datasets. The four species considered were: \textit{Magnolia grandiflora} (magnolia), \textit{Fraxinus excelsior} (ash), \textit{Betula} spp$.$ (birch), and \textit{Acer pseudoplatanus} (sycamore). The tree species are henceforth referred to by their common names. Table \ref{tab:jackson_trees} lists the approximate height, $h$, diameter at breast height, DBH, and elastic modulus, $E$, for each of the 13 test trees. Tree ID numbers are also listed for the subset of trees from the \citet{Jackson2018StrainUK} dataset for reference. Approximate values of $E$ were obtained from literature-reported values \citep{Niklas2010WorldwideDensity, Green1999MechanicalWood}.

\begin{figure}[!h]
    \includegraphics[width=\linewidth]{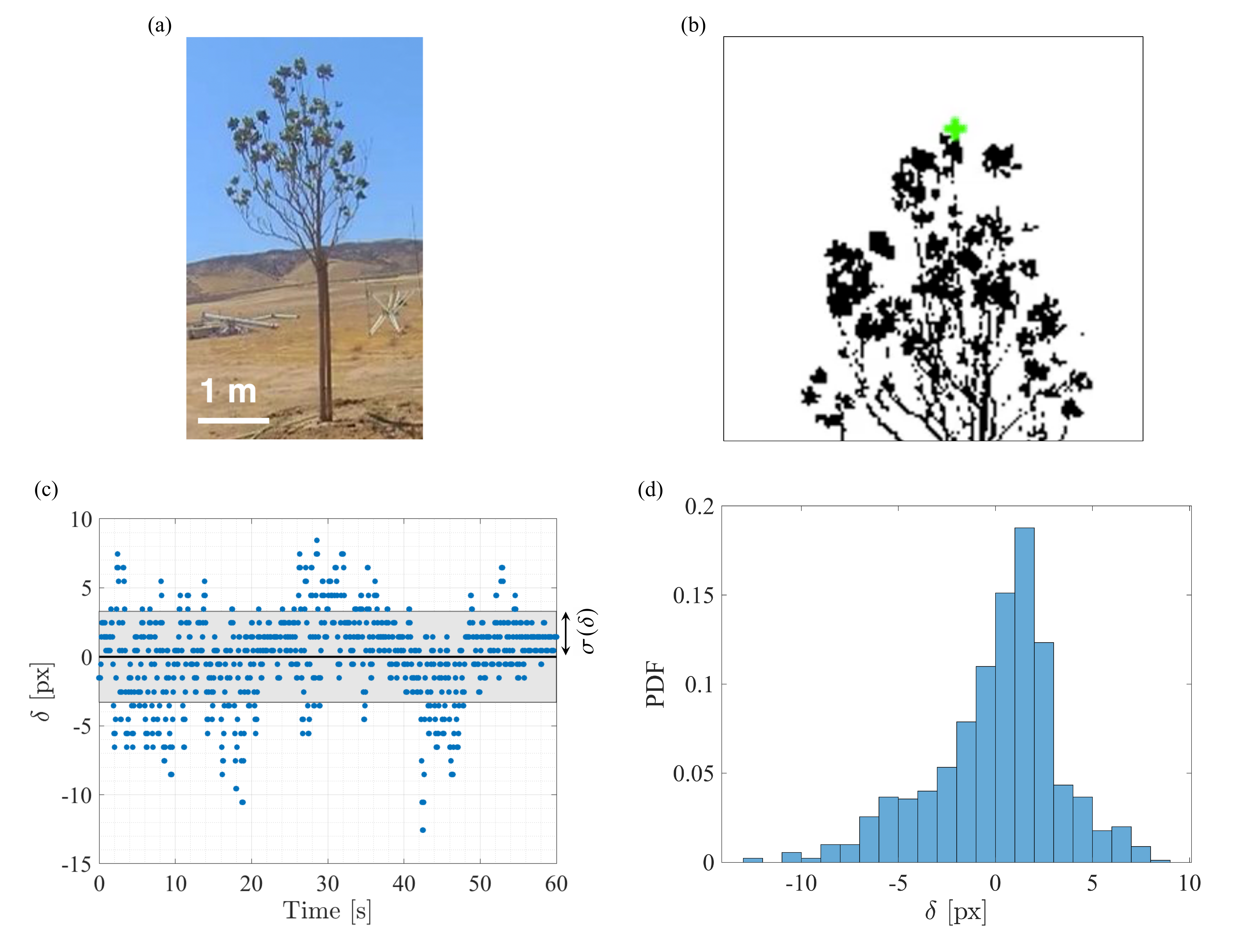}
    \centering
    \caption{(a) The \textit{Magnolia grandiflora} specimen used for analysis. (b) An example of a cropped and binarized image of the treetop. The top-most point on the tree (marked here with the green `+') was located in each frame for tracking in order to measure $\delta(t)$. (c) Plot of $\delta$ vs. time for a representative 1-minute video clip ($\bar{U} = 9.17$ ms$^{-1}$). The gray band shows $\pm \sigma(\delta)$. As noted in section \ref{sec:derivation}, $\sigma(\delta)$ was used to quantify the sway amplitude for these video experiments; (d) Probability density function (PDF) of strain measurements taken during the averaging window.}
    \label{fig:binarized}
\end{figure}

\subsubsection{Direct field measurements}
\label{sec:flora_methods}
Observations of the dynamic sway of a magnolia tree were captured in 1-minute video clips recorded in an open field with flat terrain in Lancaster, California in the United States. This test tree was the same specimen that was used by \citet{Cardona2019SeeingNetwork} to train a machine learning model. A photo of the tree is shown in figure \ref{fig:binarized}a, and tree properties are listed in table \ref{tab:jackson_trees}. The data were collected during August, 2018, during daylight hours when the lighting conditions allowed for the treetop to be easily tracked. Videos were recorded at 15 frames per second (fps). A 150 $\times$ 150 pixel (px) region of interest capturing the top of the tree was used for analysis. To measure the deflection, the frames were binarized to distinguish the tree from the background, and the top-most pixel of the tree was tracked over time. The deflection, $\delta(t)$, was measured by calculating the displacement in the horizontal direction in each frame relative to the mean position over the 1-minute period (900 frames). Deflection was measured at integer pixel resolution. An example of a binarized frame with the treetop detected is shown in figure \ref{fig:binarized}b, along with a timetrace of $\delta(t)$ over a 1-minute averaging window (figure \ref{fig:binarized}c).

 Wind speeds were recorded with an anemometer on site (Thies First Class) positioned at 10 m height and located approximately 60 m from the tree. In selecting time periods to analyze, the incident wind direction was fixed at 250$^\circ \pm 10^\circ$, which is approximately 50$^\circ$ from the plane of the recorded images. This allowed for video clips to be compared without correcting for changes in incident wind direction with respect to the camera angle. The turbulence intensity varied from 10 to 12\% during experiments, consistent with an assumed approximately constant value of $I_u$ in equation \ref{eq:dynamic_model}.

\subsubsection{Indirect field measurements}
\label{sec:sg_methods}


The dataset collected by \citet{Jackson2018StrainUK} comprised strain gage data for 21 broadleaf trees located in Wytham Woods, a broadleaf forest in southern England. Each tree was instrumented with a pair of perpendicular strain gages installed at 1.3 m height on the trunk. Strain gage data were recorded at 4 Hz. The dataset also included wind data collected in a walkway within the forest canopy with a cup anemometer at a time resolution of 0.1 Hz. The present analysis focuses on a subset of data collected during winter months (January-February, 2016) corresponding with the absence of foliage for these species of broadleaf trees as noted by \citet{Jackson2019FiniteData}. The subset of 12 trees considered in this work are listed in table \ref{tab:jackson_trees} along with their approximate dimensions.

The model given in equation \ref{eq:dynamic_model} is based on deflection, $\delta$, but it can be equivalently applied to the bending strain, $\varepsilon$, measured from the strain gages. Bending strain in an idealized cantilever beam is given by:

\begin{align}
    \varepsilon = \frac{Mc}{EI} \label{eqn}
\end{align}

\noindent where $M$ is the bending moment, $c$ is the distance from the neutral axis, $E$ is the elastic modulus, and $I$ is the area moment of inertia. For a cantilever beam subject to a uniform distributed load, the bending moment, $M$ is given by:

\begin{align}
    M = \frac{fx^2}{2}
\end{align}

\noindent where $f$ is the force per unit length and $x$ is the position along the length of the beam. The tip deflection is given by:

\begin{align}
    \delta = \frac{fL^4}{8EI}
\end{align}

The strain, $\varepsilon$, and tip deflection, $\delta$, both scale with the force per unit length, $f$. Thus, the strain is proportional to the tip deflection ($\varepsilon \propto \delta$) for a given loading, and the strain gage measurements can be used directly in the model in place of the tip deflections. The standard deviation of $\varepsilon$ is used to define the sway amplitude for analyses of the strain gage dataset, and in this case the model relating sway amplitude and wind speed is:

\begin{align}
    \sigma(\varepsilon) &\propto I_u \bar{U}^2
    \label{eq:dynamic_model_epsilon}
\end{align}

\begin{figure}[hbt!]
    \centering
    \includegraphics[width=\linewidth]{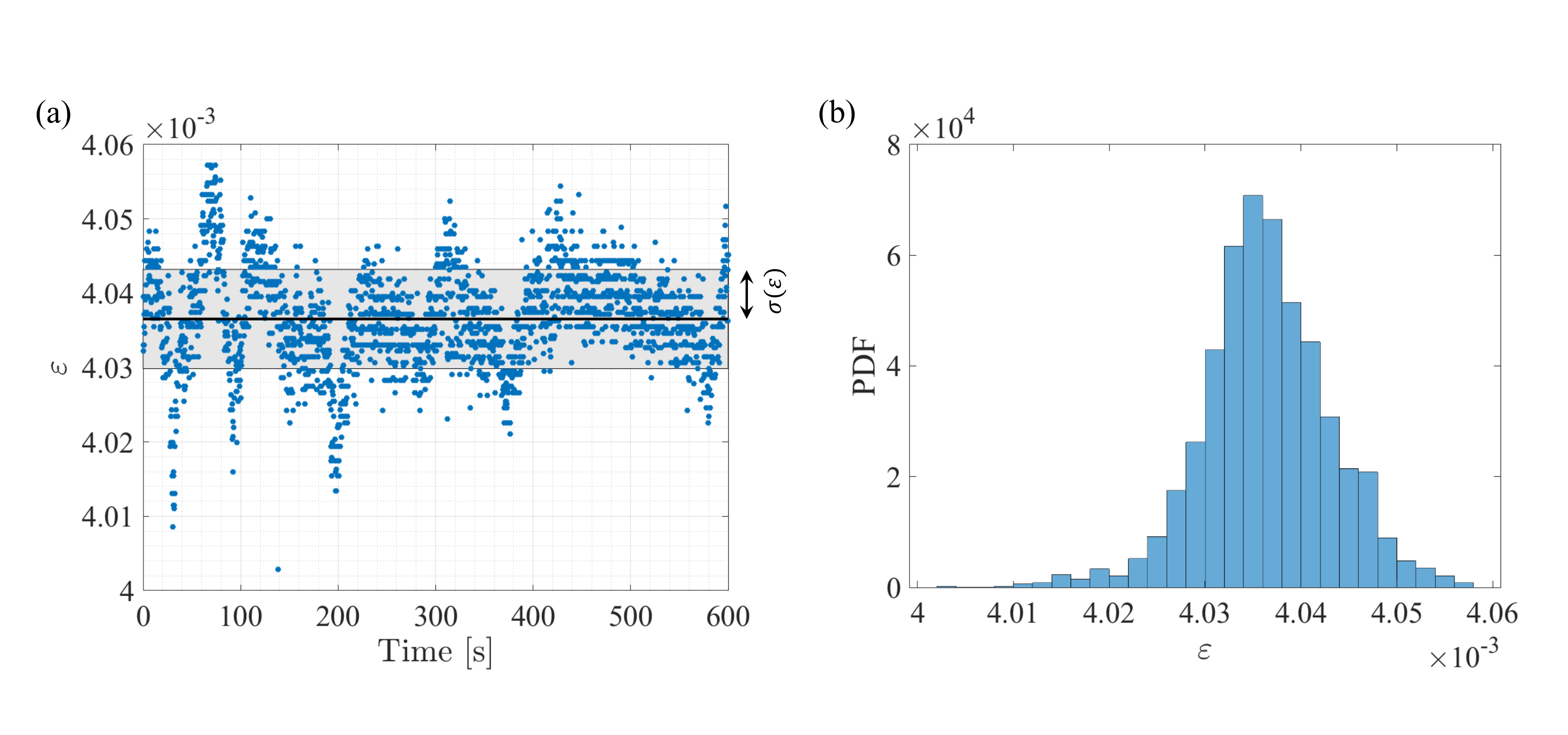}
    \centering
    \caption{Representative example of strain measurements over a 10-minute averaging window for a sycamore tree (tree \# 17) with $\bar{U}=1.08$ ms$^{-1}$. (a) Strain vs. time with the mean value shown with the black line and $\pm \sigma(\varepsilon)$ shown with the gray band; (b) Probability density function (PDF) of strain measurements taken during the averaging window.}
    \label{fig:example_timetrace_pdf}
\end{figure}

To compare experimental results to the physical model, sway amplitudes, mean wind speeds, and turbulence intensities were quantified from the \citet{Jackson2018StrainUK} dataset over 10-minute averaging periods. Note that this averaging period is longer than the 1-minute period used for the video dataset (section \ref{sec:flora_methods}). This longer averaging period was afforded by the abundance of available data. We observed improved model agreement using the longer, 10-minute averaging windows (this is discussed further in the supplementary material, section \ref{sec:window_length}). Timestamps were matched to retrieve samples for which both anemometer-recorded wind data and strain gage data were available over the course of the full 10-minute averaging periods. The averaging periods did not overlap (i.e. each sample represented a unique window in time). Values of $\bar{U}$ and $I_u$ were calculated from the instantaneous measurements of $U(t)$ for each averaging period. The strain, $\varepsilon(t)$, was calculated as:

\begin{align}
    \varepsilon(t) = \sqrt{\varepsilon_N^2(t) + \varepsilon_E^2(t)}
\end{align}

\noindent where $\varepsilon_N$ and $\varepsilon_E$ are the strain measurements recorded by the northward and eastward oriented strain gages comprising a perpendicular pair for a given tree. The instantaneous strain measurements, $\varepsilon(t)$, were used to calculate the sway amplitude, $\sigma(\varepsilon)$, for each averaging window. A representative example timetrace of the strain over a 10-minute averaging window is shown in figure \ref{fig:example_timetrace_pdf}a. The probability density function (PDF) of strain measurements over the averaging window is also shown (figure \ref{fig:example_timetrace_pdf}b).

\subsection{Tree response regimes}
\label{sec:regimes}
The response of a tree subject to incident wind loading depends on both fluid and structural properties. The Cauchy number, $Ca$, and slenderness ratio, $S$, are useful in determining whether a tree is likely to show large static bending \citep{deLangre2008EffectsPlants}. The parameters $Ca$ and $S$ are defined as:

\begin{align}
    Ca &= \frac{\rho U^2}{E} \label{ca}\\
    S &= \frac{L}{l}\label{s}
\end{align}

\noindent where $L$ and $l$ are the maximum and minimum dimensional lengths of the cross section perpendicular to the flow direction respectively. Large static deformation is expected when $Ca S^3 > 1$ \citep{deLangre2008EffectsPlants}. Therefore, less slender trees will resist large observable static bending until higher wind speeds are reached. Values of $CaS^3$ were approximated for the trees analyzed in this work. The lengths, $L$ and $l$, were taken to be the tree height ($h$) and the diameter at breast height (DBH) respectively. The trees of interest in this study correspond to values of $CaS^3 \leq 5 \times 10^{-3}$, suggesting that large static deformations are not present. This limits the applicability of the mean bending method of \citet{Cardona2021WindInteractions}.


\begin{figure}[!h]
    \includegraphics[width=0.7\linewidth]{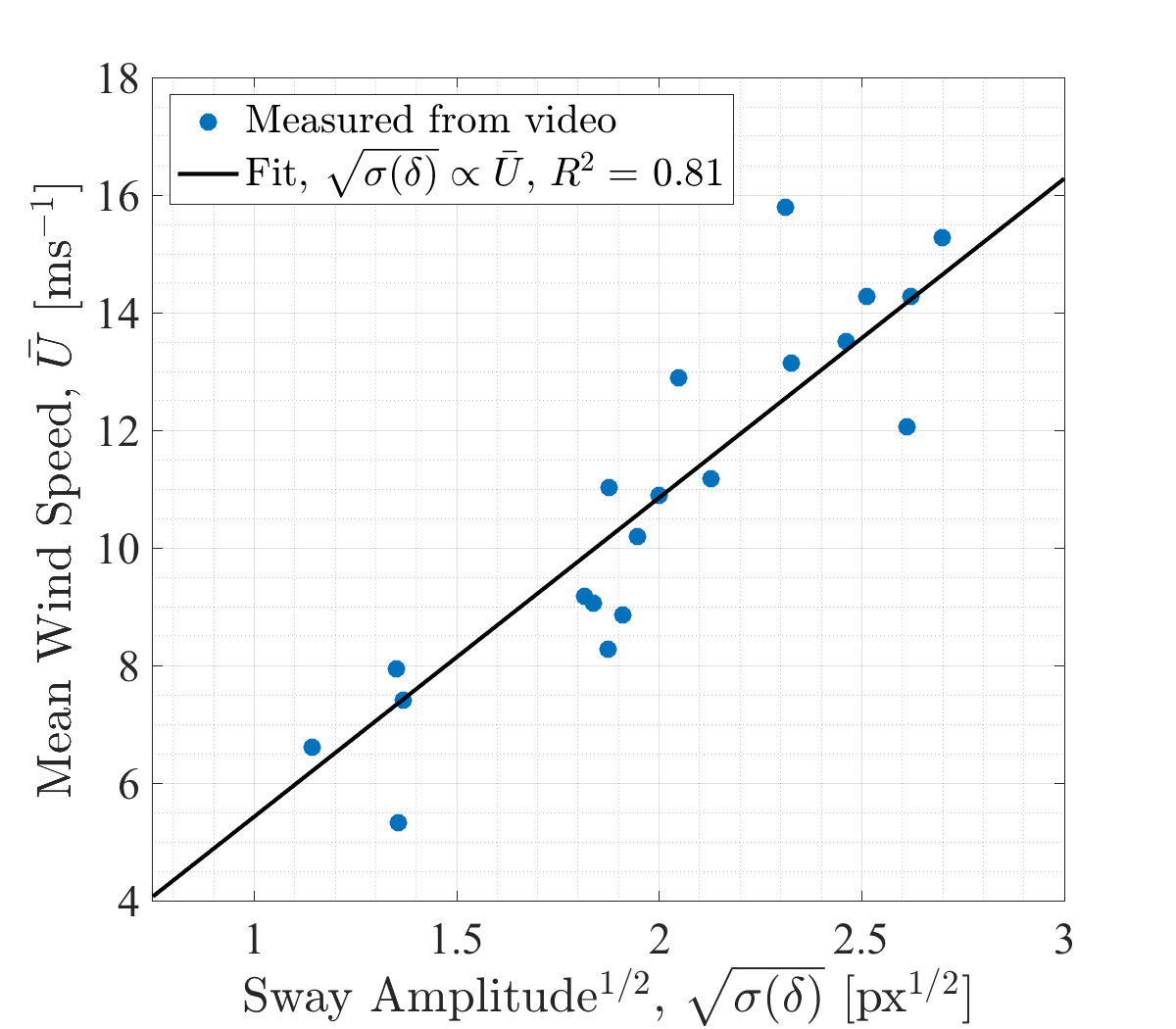}
    \centering
    \caption{Mean wind speed, $\bar{U}$, vs. the square root of sway amplitude measured from video frames. Black lines indicate best-fit for proportional model with $R^2$ value of 0.81.}
    \label{fig:flora}
\end{figure}

\section{Results}

\subsection{Model comparison to direct field measurements from video data}
\label{sec:flora_results}

Figure \ref{fig:flora} shows the mean wind speed plotted against the square root of the sway amplitude for the field measurements of the magnolia tree from the video dataset. Datapoints are shown for each of the 1-minute video clips captured in the field experiments. The best-fit line assuming proportionality (i.e. $\sqrt{\sigma(\delta)} \propto \bar{U}$) is shown with the black line (calculated using ordinary least squares). The experimental data agree well with the proportional model, with the best-fit line yielding $R^2 = 0.81$. These results demonstrate that the proposed physical model (i.e. equation \ref{eq:dynamic_model}) characterizes the relationship between tree sway and mean wind speed captured in this video dataset.

\subsection{Model comparison to indirect field measurements from strain gage data}
\label{sec:results_sg}

\begin{figure}[hp!]
    \centering
    \includegraphics[width=\linewidth]{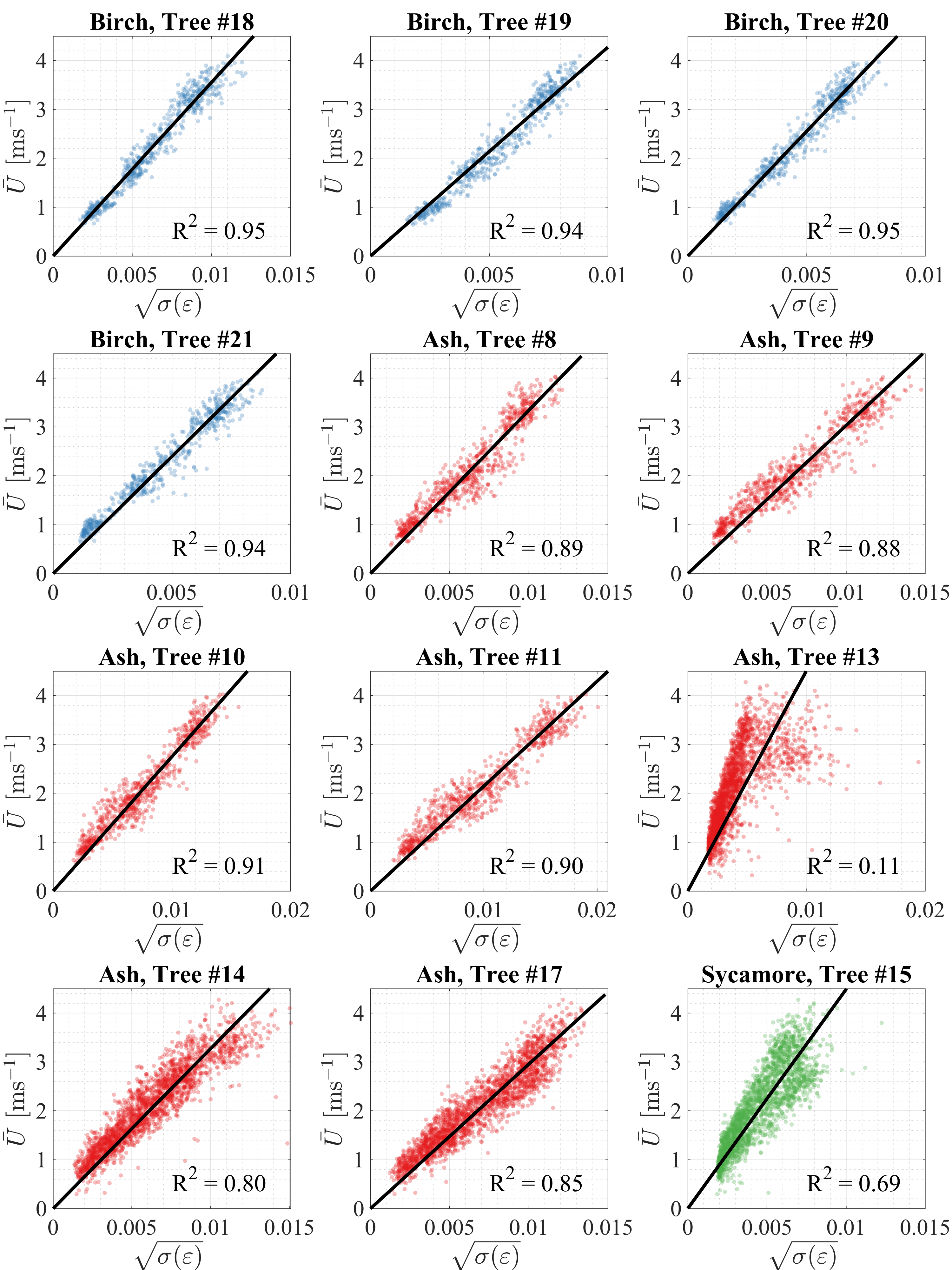}
    \caption{Mean wind speed, $\bar{U}$, vs. the square root of sway amplitude measured from strain gages for 12 trees. Black lines indicate best fit for proportional model with $R^2$ values as shown. Representatives from three tree species are shown: birch (blue), ash (red), and sycamore (green). Agreement is generally good except for outlier tree $\#$13.}
    \label{fig:external_data}
\end{figure}

Figure \ref{fig:external_data} shows the mean wind speed versus the square root of the sway amplitude for each of the 12 trees analyzed from the \citet{Jackson2018StrainUK} dataset. In the engineering application employing the physical model for anemometry, both $\bar{U}$ and $I_u$ would be unknown, and the oscillatory motion of the structure would be the sole experimentally measured quantity. Therefore, $I_u$ was not considered to generate these results (i.e. $I_u$ assumed constant). Model sensitivity to $I_u$ is further discussed in the supplementary material, section \ref{sec:TI}. The best-fit line assuming the proportional relationship $\bar{U} \propto \sqrt{\sigma(\varepsilon)}$ is shown for each tree (again calculated using ordinary least squares). This proportional relationship characterizes the trend in the experimental data well, with clear agreement for all trees with the exception of one outlier specimen (tree \#13). 

\section{Discussion}
\subsection{Application to anemometry and calibration reference point considerations}
\label{sec:ref_considerations}
The results shown in figures \ref{fig:flora} and \ref{fig:external_data} suggest that the proposed model captures the trend in relating structural sway amplitude to the mean wind speed. This is apparent in the clear agreement between the best-fit proportional lines and experimental sway measurements for each of the various sample trees. As discussed in section \ref{sec:intro}, the motivation behind this modelling effort is ultimately to use observations of the structural sway for anemometry. If the proportionality constant relating $\sqrt{\sigma(\delta))}$ and $\bar{U}$ was known \textit{a priori}, then the relationship could be applied directly to estimate $\bar{U}$ given $\delta(t)$ (or $\varepsilon(t)$ in the case of strain gage measurements). However, the proportionality constant is inherently structure-specific, dependent upon unique structure geometries and material properties. In order to determine a structure-specific proportionality constant (i.e. the slope of the best-fit line), a measurement campaign capturing both $\bar{U}$ and $\delta(t)$ would be necessary. This would require an anemometer to be installed over a long time duration to capture the structure response under varying wind loads, which would undermine the purpose of using the structure itself as an anemometer.

\begin{figure}[b!]
    \centering
    \includegraphics[width=\linewidth]{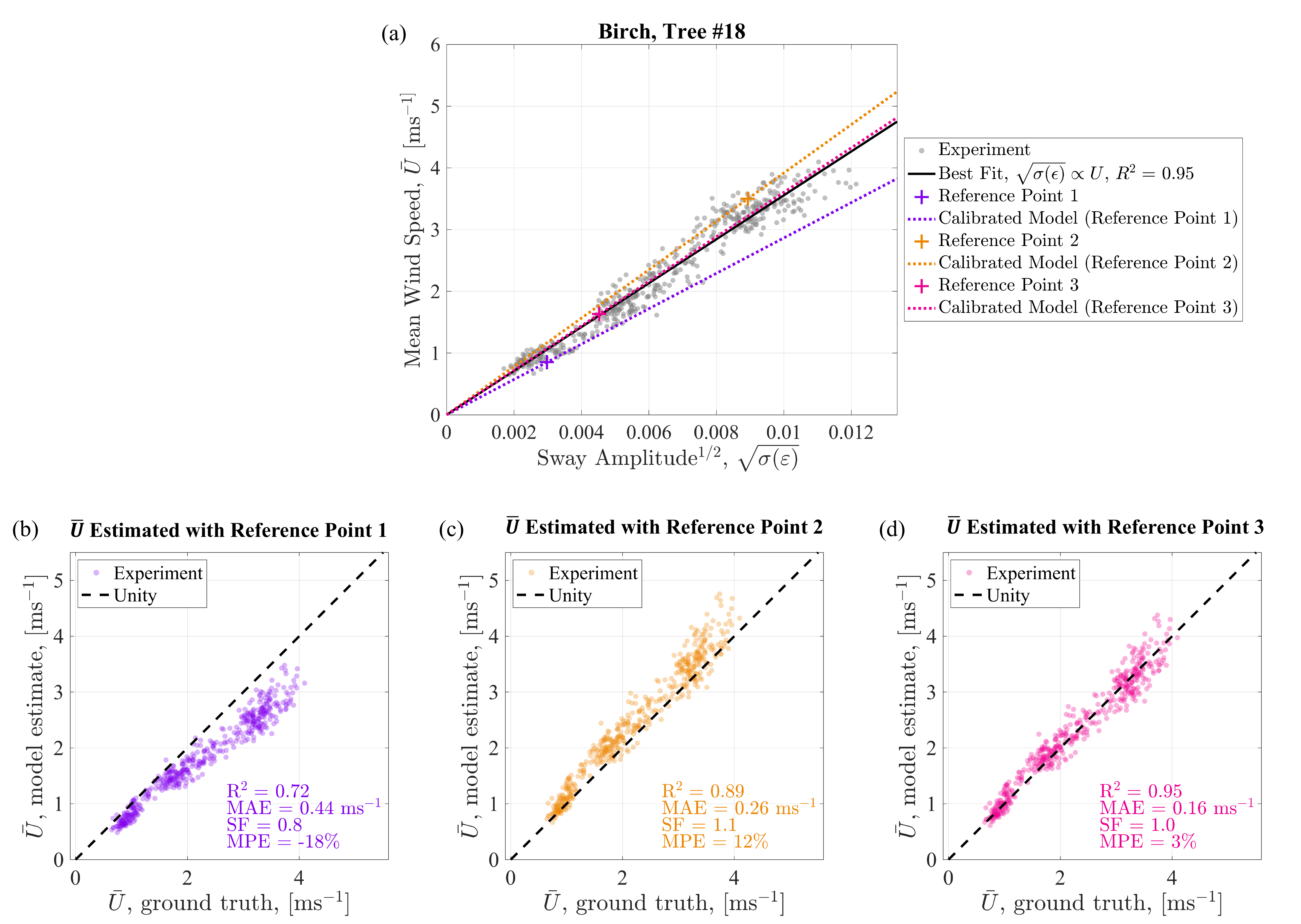}
    \caption{Example showing reference-calibrated model performance for a birch tree. (a) Mean wind speed, $\bar{U}$, vs. the square root of sway amplitude for representative birch tree (tree \#18). The best-fit line is shown in black. Three example reference points are shown with `+' marks, and the resulting reference-calibrated models are shown with the dotted lines. (b-d) Model-estimated $\bar{U}$ vs. ground truth for reference-calibrated models using reference points 1, 2, and 3 respectively. Evaluation metrics including $R^2$, MAE, SF, and MPE are shown.}
    \label{fig:birch_example}
\end{figure}

One possibility for using the proposed anemometry method without an extensive measurement campaign is to calibrate the model using wind speed and structural sway measurements for a single averaging period. This allows for the inference of normalized ratios of wind speeds at the specific site, and the dimensional wind speeds can be recovered if the reference wind speed is known. This approach was taken in the visual anemometry method using mean deflections proposed by \citet{Cardona2021WindInteractions}. The trade-off is that the estimated model slope for a given structure will be based on a single calibration point, and may not capture the trend as well. The slope of the reference-calibrated model is $\overline{U_0}/\sqrt{\sigma(\varepsilon_0)}$, where $\overline{U_0}$ and $\sqrt{\sigma(\varepsilon_0)}$ are quantities measured over the reference time period. Several metrics are used below to evaluate the goodness-of-fit of these reference-calibrated models, including the $R^2$ value, the mean absolute error (MAE), the scale factor between the reference-calibrated and best-fit slopes (SF), and the mean percentage error (MPE).

An illustrative example of this calibration reference point approach is shown in figure \ref{fig:birch_example}. Figure \ref{fig:birch_example}a shows a plot of $\bar{U}$ vs. $\sqrt{\sigma(\varepsilon)}$ for a birch tree. Three hypothetical reference points are shown by `+' marks, with the resulting calibrated models shown with the dotted lines. Figures \ref{fig:birch_example}b-d show how the three reference-calibrated models perform when applied to estimate $\bar{U}$ from the tree sway measurements. Plots show the model-estimated wind speed versus the ground truth wind speed for each of the three reference-calibrated models. The dashed black lines indicate a perfect one-to-one relationship. Figures \ref{fig:birch_example}b-d demonstrate how reference-calibrated model performance depends on the chosen reference point. The reference-calibrated models will tend to systematically over-estimate or under-estimate $\bar{U}$ depending on whether the calibrated slope is greater than or less than the best-fit slope. For example, in figure \ref{fig:birch_example}a, Reference Point 1 lies below the best fit line (i.e. the sway amplitude was higher than usual for the incident wind speed), which resulted in a model that systematically underestimated wind speeds compared to ground truth (figure \ref{fig:birch_example}b). In contrast, Reference Point 2 lies above the best fit line (figure \ref{fig:birch_example}a), resulting in a model that systematically overestimated wind speeds compared to ground truth (figure \ref{fig:birch_example}c). Reference points that lie close to the best fit line (e.g. Reference Point 3) yield the best results (figure \ref{fig:birch_example}d).


\begin{figure}[hbt!]
\centering
    \includegraphics[width=0.9\linewidth]{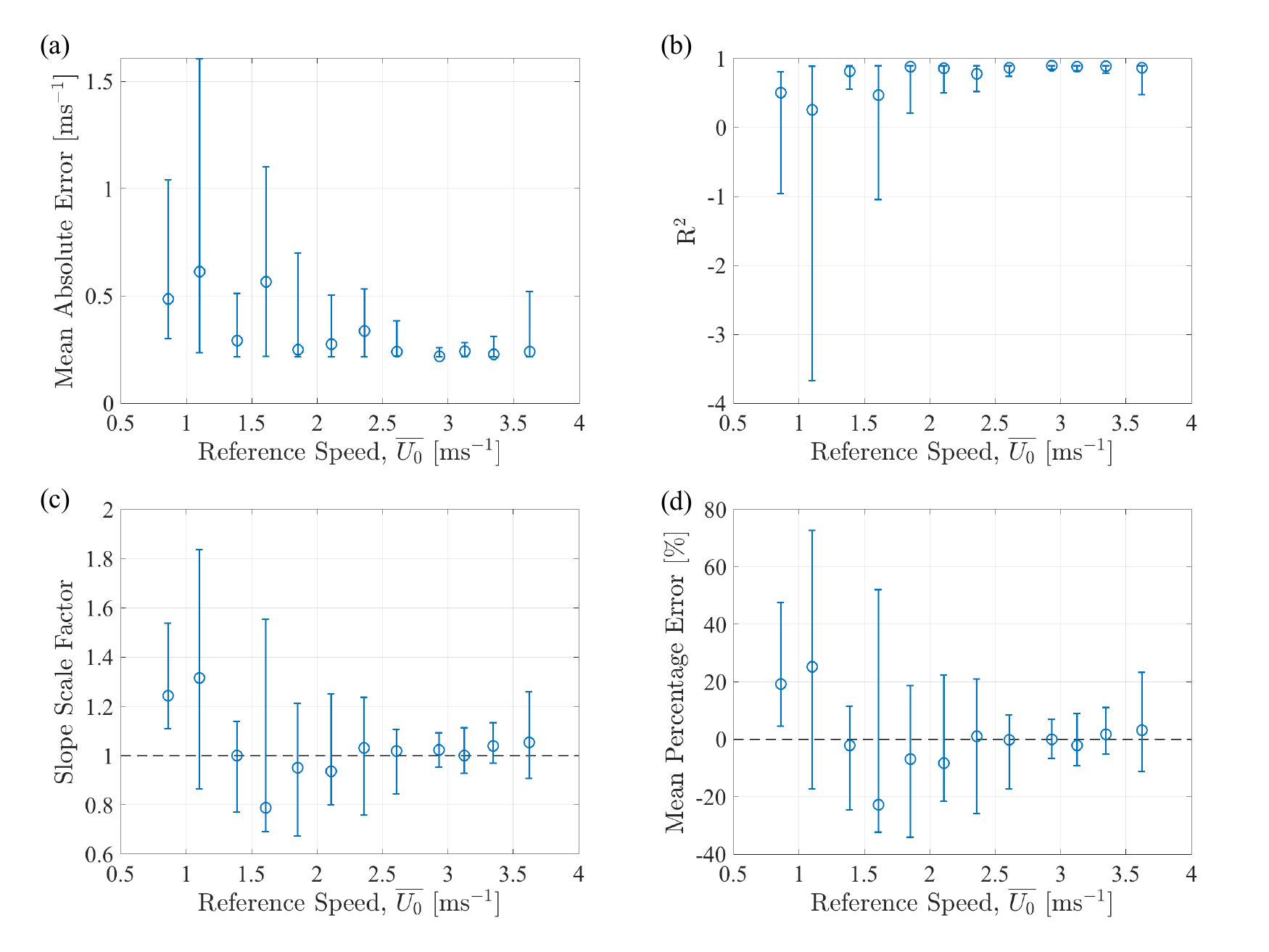}
    \caption{Model assessment metrics vs. reference wind speed $\bar{U_0}$ for all 12 strain gaged trees: (a) mean absolute error (MAE); (b) $R^2$; (c) slope scale factor (SF); and (d) mean percentage error (MPE). Markers represent the median values, and error bars denote the 5\textsuperscript{th} and 95\textsuperscript{th} percentile values.}
    \label{fig:error_metrics}
\end{figure}

A reference condition with a higher flow speed may be beneficial because of the higher signal-to-noise ratio in measuring the sway amplitude. A deviation of the reference point from the best-line at a low reference wind speed will lead to a greater discrepancy in the slope compared to a reference point taken at a higher wind speed with the same deviation. To further illustrate this point, reference-calibrated models were evaluated as a function of reference wind speed. For each of the 12 strain gaged trees, the available samples were sorted by $\bar{U}$ in bins of 0.25 ms$^{-1}$. For each bin with at least 20 samples for a given tree, 10 samples were chosen at random and held out as possible reference conditions. The remaining samples were used as test conditions. Each reference condition was used to calibrate a model to estimate the $\bar{U}$ for the test conditions. The error metrics were calculated for each reference-calibrated model. Figure \ref{fig:error_metrics} shows error metrics as functions of $\overline{U_0}$ for all trees combined. The reference-calibrated models tend to improve with higher $\overline{U_0}$. The model performance also becomes more consistent, which is demonstrated by the decreasing size of the error bars marking the 5\textsuperscript{th} and 95\textsuperscript{th} percentiles.

\section{Conclusion}
In the present work, a physical model was developed relating the mean wind speed, $\bar{U}$, to the amplitude of structural oscillations, $\sigma(\delta)$, in response to incident wind. The model was compared to two experimental field datasets with trees as the objects of interest. The first was a video dataset capturing the swaying of a magnolia tree, and the second was a subset of the publicly available data collected by \citet{Jackson2018StrainUK} comprising strain gage data of various trees in a broadleaf forest. Between these two datasets, 13 trees were analyzed representing 4 different species. The physical model agreed well with the experimental measurements from both datasets. The relationship $\bar{U} \propto \sqrt{\sigma(\delta)}$ was robust for the trees over the range of conditions analyzed here. However, further consideration should be given to the effect of large changes in turbulence intensity, $I_u$, especially at high wind speeds.

The excellent agreement between the model and experimental results suggests that the model can be used towards visual anemometry, where structural sway recorded in video data can be used to measure wind speeds. However, the model scaling is structure-specific, so model application to anemometry requires a calibration for each structure of interest. Alternatively, the method can be used to infer the normalized ratios of wind speeds present at the site of interest. The proposed visual anemometry method based on sway amplitude provides advantages over the previously developed technique by \citet{Cardona2021WindInteractions}, because it can be used on large trees that may not exhibit noticeable mean bending, and it does not require additional calibration measurements to be taken in the absence of wind.


\begin{Backmatter}


\paragraph{Funding Statement}
This work was supported by the National Science Foundation (grant CBET-2019712).

\paragraph{Competing Interests}
The authors report no conflict of interest.

\paragraph{Data Availability Statement}
The data used in this work will be made available upon request. 

\paragraph{Author Contributions}
Conceptualization: JLC; JOD. Methodology: JLC; JOD. Investigation: JLC. Software: JLC. Data analysis: JLC; JOD. Funding acquisition: JOD.

\paragraph{Supplementary Material}
Additional information can be found in the supplementary material.

\bibliographystyle{apalike}
\bibliography{references.bib}

\end{Backmatter}

\newcommand{\beginsupplement}{%
        \setcounter{table}{0}
        \renewcommand{\thetable}{S\arabic{table}}%
        \setcounter{figure}{0}
        \renewcommand{\thefigure}{S\arabic{figure}}
                \setcounter{section}{0}
        \renewcommand{\thesection}{S\arabic{section}}
        \setcounter{page}{0}
        \renewcommand{\thepage}{S\arabic{page}}
     }
     \newpage
\beginsupplement


\noindent {\color{jourcolor} \title{\Large \textbf{Supplementary material}}}

\vspace{1em}
\noindent Jennifer L. Cardona and John O. Dabiri

\section{Model sensitivity to turbulence intensity, sway amplitude definition, and averaging period}
\label{sec:sensitivity}

The results presented in section \ref{sec:results_sg} assumed constant $I_u$ in order to directly relate $\bar{U}$ and the structural sway, since $I_u$ would be unknown in most engineering applications of the presently proposed anemometry method. However, the derived physical model (equation \ref{eq:dynamic_model_epsilon}) suggests that sway amplitude should depend on $I_u$. Section \ref{sec:TI} discusses the results of including the variable $I_u$. Model sensitivity is also evaluated with respect to other parameters including the definition of sway amplitude (section \ref{sec:mad}), and the time averaging window (section \ref{sec:window_length}). A summary of these results is given in table \ref{tab:sensitivity} and is described in section \ref{sec:summary_sensitivity}.

\subsection{Incorporating turbulence intensity}
\label{sec:TI}
\begin{figure}[hbt!]
    \centering
    \includegraphics[width=\linewidth]{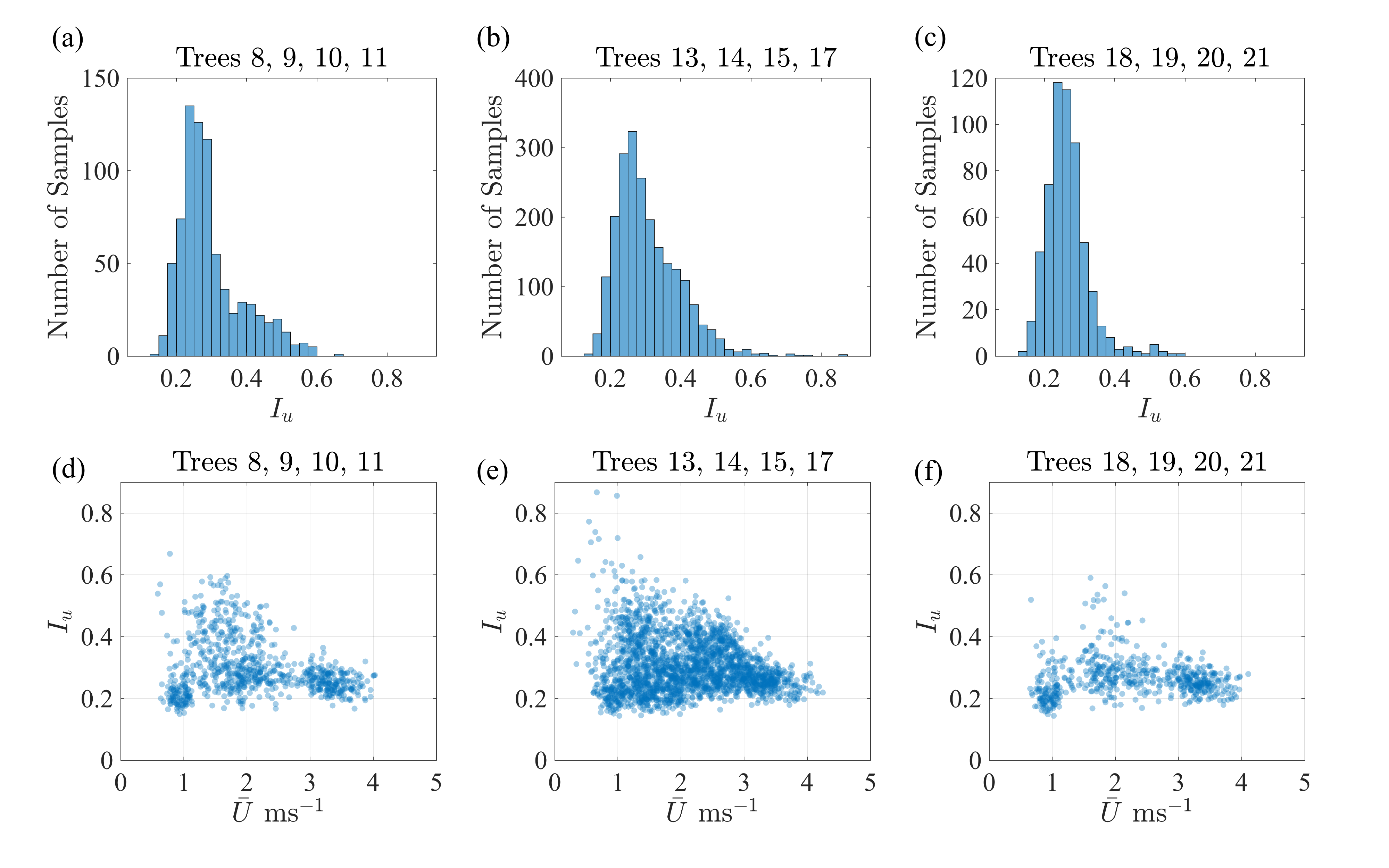}
    \caption{(a-c) Histograms of $I_u$ and (d-f) plots of $I_u$ vs. $\bar{U}$ for trees grouped by data logger (i.e. separate panels show trees in subsets [8, 9, 10, 11], [13, 14, 15, 17], and [18, 19, 20, 21]).}    
    \label{fig:TI_distributions}
\end{figure}

Figure \ref{fig:TI_distributions} shows the distributions of $I_u$ (a-c), and plots of $I_u$ versus $\bar{U}$ (d-f) for the trees in each of the three data logger groups. The most common value of turbulence intensity in the forest was approximately 25\%, and it decreased with increasing wind speed. These trends are expected in comparison with the lower turbulence intensity measured for higher winds in the open field. Nonetheless the scaling predicted by the model in equation \ref{eq:dynamic_model} which assumes $u' << \bar{U}$ remains effective when applied to the forest data. To analyze the effects of a non-constant turbulence intensity, $I_u$ was allowed to remain variable in equation \ref{eq:dynamic_model_epsilon} ($\bar{U} \propto \sqrt{\sigma(\varepsilon)/I_u}$). Figure \ref{fig:sensitivity_comparison}b shows the results for a representative tree (tree \#18) with the known values of $I_u$ incorporated. These results can be compared to the baseline case assuming constant $I_u$ (figure \ref{fig:sensitivity_comparison}a). $R^2$ values are reported for all trees in table \ref{tab:sensitivity}.

Allowing for variable $I_u$ did not appear to improve model agreement compared to the baseline case. One possible explanation is offered by the range of wind speeds for these experiments, which is relatively low ($\bar{U} < 4$ ms$^{-1}$). This means that fluctuations have small magnitudes in dimensional terms and therefore a smaller effect on the tree structure dynamics. The higher values of $I_u$ occurred at lower wind speeds (figure \ref{fig:TI_distributions}d-f), which would further contribute to this effect. Prior work also suggests that gusts at low wind speeds have little effect on tree sway for trees within forest canopies because of a lack of gust penetration into the canopy \citep{Gardiner1994WindForest, Gardiner1997FieldStability}.


\subsection{Median absolute deviation as an alternative amplitude definition}
\label{sec:mad}

As discussed in section \ref{sec:derivation}, the standard deviation was chosen as a measure to represent the sway amplitude for the trees analyzed in this study. However, deflection or strain measurements over an averaging period do not follow perfectly normal distributions (figures \ref{fig:binarized}d and \ref{fig:example_timetrace_pdf}b). It has been suggested that the median absolute deviation (MAD) may a be more robust measure of dispersion for non-normal data distributions, since it is less sensitive to long tails than $\sigma$ \citep{Ruppert2010StatisticsEngineering}. The median absolute deviation is calculated as:

\begin{align}
    \text{MAD} = \text{median}(|X_i - \text{median}(X)|)
\end{align}

\noindent where $X_i$ are the population samples. MAD($\varepsilon$) was applied to the strain gage measurements as an alternative measure of sway amplitude. Model agreement was robust to the choice of measure, as demonstrated in figure \ref{fig:sensitivity_comparison}c and table \ref{tab:sensitivity}, which show that results using MAD were very similar to the baseline case.

\subsection{Temporal averaging period length}
\label{sec:window_length}
The temporal averaging period is both a practical consideration in terms of data collection and processing time, as well as an important factor based on the potential applications. For the video dataset analysis described in section \ref{sec:flora_methods}, a 1-minute averaging window was used, and model agreement was observed (figure \ref{fig:flora}). The abundance of data available in the \citet{Jackson2018StrainUK} dataset allowed for the selection of a longer time averaging period (10-minute periods were used in the baseline results shown in figures \ref{fig:external_data} and \ref{fig:sensitivity_comparison}). The results from the application of 1-minute averaging windows are shown in figure \ref{fig:sensitivity_comparison}d and table \ref{tab:sensitivity}. The longer 10-minute averaging periods led to better agreement than 1-minute averaging periods. Model agreement with the shorter 1-minute averaging periods may have suffered due to the spatial separation between the anemometer used to measure $\bar{U}$ and the structures of interest (for many of the trees, it is located approximately 230 m away as detailed by \citet{Jackson2019FiniteData}). The trees were also located within a forest, where hyper-local conditions may be spatially variable, especially over shorter timescales. 

The choice of a 10-minute averaging window is still an appropriate averaging window to provide useful measurements of mean wind speed in the context of engineering applications. For instance, 10-minute averaging windows are common in wind speed measurements for wind energy applications \citep{Mathew2006WindEconomics}.

\subsection{Summary of results from model parameter changes}
\label{sec:summary_sensitivity}
\begin{figure}[hbt!]
    \centering
    \includegraphics[width=\linewidth]{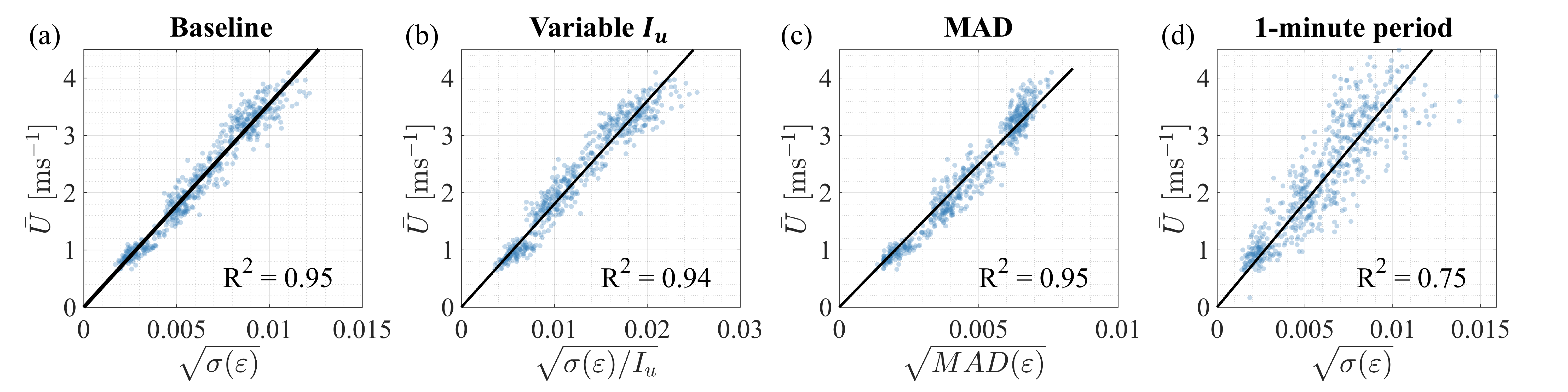}
    \caption{Representative example showing experimental results compared to model relationship for tree \#18 considering (a) the baseline case (constant $I_u$, $\sigma$ defining sway amplitude, and 10-minute averaging windows); (b) the case allowing for variable $I_u$; (c) the case using median absolute deviation (MAD) to define sway amplitude instead of $\sigma$; and (d) the case using 1-minute time averaging periods instead of 10-minute averaging periods.}    
    \label{fig:sensitivity_comparison}
\end{figure}

The results of adjusting model parameter choices are summarized in table \ref{tab:sensitivity}, and a representative example is shown for tree \#18 in figure \ref{fig:sensitivity_comparison}. Including the variable turbulence intensity, $I_u$, did not appear to have a large impact on the results. However, as discussed in section \ref{sec:TI}, the range of mean wind speeds (and hence, the magnitude of fluctuations) was relatively low, which may have made effects difficult to detect in the swaying of trees of this size. The effect of $I_u$ should be further considered in future studies, especially in cases where there is a substantial range of $I_u$ at higher wind speeds. Model performance was also consistent with the baseline case when median absolute deviation (MAD) was used as an alternative to standard deviation in characterizing the sway amplitude. Model agreement was sensitive to the choice of time averaging period. The 10-minute averaging period proved to be a better choice compared to shorter 1-minute averaging periods in this case.

\begin{table*}[!hbt]
  \begin{center}
\caption{$R^2$ values for best-fit line compared to experimental data for each tree. Results are reported for the model applied in the baseline case (constant $I_u$, $\sigma$ defining sway amplitude, and 10-minute averaging periods), the case allowing for variable $I_u$, the case using median absolute deviation (MAD) to define sway amplitude, and the case using 1-minute time averaging period instead of 10-minute averaging period.}
\vspace{12 pt}
  \begin{tabular}{lc|cccc}
    
    & & $R^2$ & $R^2$ & $R^2$ & $R^2$\\
    Tree ID \# & Species & (Baseline) & (Variable $I_u$) & (MAD) & (1-min. avg. period)\\
    \midrule
      8 &  Ash & 0.89 & 0.90 & 0.90 & 0.56\\
      \hline
      9 & Ash & 0.88 & 0.83 & 0.90 & 0.47\\
      \hline
      10 & Ash & 0.91 & 0.87 & 0.91 & 0.56\\
      \hline
      11 & Ash & 0.90 & 0.87 & 0.90 & 0.53\\
      \hline
      13 & Ash & 0.11 & 0.24 & 0.09 & 0.18\\
      \hline
      14 & Ash & 0.80 & 0.71 & 0.81 & 0.43\\
      \hline
      15 & Sycamore & 0.69 & 0.83 & 0.74 & 0.37\\
      \hline
      17 & Ash & 0.85 & 0.89 & 0.86 & 0.47\\
      \hline
      18 & Birch & 0.95 & 0.94 & 0.95 & 0.75\\
      \hline
      19 & Birch & 0.94 & 0.91 & 0.94 & 0.79\\
      \hline
      20 & Birch & 0.95 & 0.95 & 0.96 & 0.74\\
      \hline
      21 & Birch & 0.94 & 0.95 & 0.96 & 0.71\\
      
  \end{tabular}
  \label{tab:sensitivity}
  \end{center}
\end{table*}

\end{document}